%% file: Paper.tex
\documentclass[final,12pt,authoryear]{elsarticle}
\makeatletter
   \def\ps@pprintTitle{%
      \let\@oddhead\@empty
      \let\@evenhead\@empty
         \def\@oddfoot{\centerline{\thepage}}%
      \let\@evenfoot\@oddfoot
   }
\makeatother
\usepackage[T1]{fontenc}
\usepackage[utf8]{inputenc}
\include{settings}

\usepackage{amssymb,amsmath,stmaryrd,footmisc}
\usepackage{esint}
\usepackage{placeins}

\usepackage{xcolor}
\usepackage{epstopdf}
\usepackage{natbib}
\usepackage{nohyperref}
\usepackage{graphicx}
 \usepackage{setspace}
\usepackage{color}
\usepackage{verbatim}
 \usepackage[usenames,dvipsnames]{pstricks}
 \usepackage{epsfig}
 \usepackage{pst-grad} 
 \usepackage{pst-plot} 
 \usepackage{fancyref}

\usepackage{etoolbox}
\usepackage{mathtools}
\preto\subequations{\ifhmode\unskip\fi}

\usepackage{physics}
\usepackage{amsmath}

\usepackage{placeins}
\usepackage{siunitx}

\usepackage[10pt]{moresize}
\usepackage{amssymb,amsmath,stmaryrd,footmisc,float}
\usepackage{color}
\usepackage{ulem}
\usepackage{graphicx}
\usepackage[makeroom]{cancel}
\usepackage{epstopdf}

\newlength{\tleft}
 \newlength{\tright}
\def\nten#1{\mathbf{#1}}

\def\dD0{\mathcal{\partial D}_0}
\def\D0{\mathcal{D}_0}

\def\dV0{\dd{V_0}}

\def\dA0{\dd{A_0}}

\usepackage{cleveref}

\begin{document}

\begin{frontmatter}

\small
\title{A Reciprocal Theorem for Finite Deformations in Incompressible Bodies}


\author[label1]{T. Henzel$^\dagger$}
\author[label2]{S. Chockalingam$^\dagger$}
\author[label1,label3]{T. Cohen\corref{cor1}}
\cortext[cor1]{Corresponding author: talco@mit.edu\\ $\dagger$ These authors contributed equally.}

\address[label1]{Massachusetts Institute of Technology, Department of Civil and Environmental Engineering, Cambridge, MA, 02139, USA}
\address[label2]{Massachusetts Institute of Technology, Department of Aeronautics and Astronautics, Cambridge, MA, 02139, USA}
\address[label3]{Massachusetts Institute of Technology, Department of Mechanical Engineering, Cambridge, MA, 02139, USA}

\begin{abstract}
 The reciprocal theorems of Maxwell and Betti are foundational in mechanics but have so far been restricted to infinitesimal deformations in elastic bodies. In this manuscript, we present a reciprocal theorem that relates solutions of a specific class of large deformation boundary value problems for  incompressible bodies; these solutions are  shown to identically satisfy the Maxwell-Betti theorem. The  theorem has several potential applications such
as development of alternative convenient experimental
setups for the study of material failure through bulk
and interfacial cavitation, and  leveraging easier numerical
implementation of equivalent auxiliary boundary
value problems. The following salient features of the theorem are noted:  (i) it applies to dynamics in addition to statics, (ii) it allows for large deformations,  (iii) generic body shapes with several potential holes, and (iv) any general type of boundary conditions. 
\end{abstract}


\end{frontmatter}


\section{Introduction}
\noindent In 1864 Maxwell reported an astonishing observation: Given an elastic body, in which we can identify two arbitrary points, $A$ and $B$; the displacement of point $A$ that results from a force applied at point $B$, is equal to the displacement that would ensue at point $B$ from the application of the same force at point $A$   \citep{maxwell1864calculation}. His \textit{reciprocal theorem}, which was later generalized and formalized by \cite{betti1872teoria}, has been foundational in structural mechanics and elasticity \citep{truesdell1963meaning,barber2002elasticity,love2013treatise, charlton1960historical}. The idea that one can experimentally observe a body deforming under a given set of boundary conditions, and then directly infer its response due to an alternative set, without solving the boundary value problem,  has proven to be extremely useful, and has become one of the most classical results in elasticity. Other than its mathematical elegance  \citep{shield1967load}, it has  been particularly useful in contact mechanics, to interpret indentation measurements conducted with different indentor shapes  \citep{garcia2018determination,managuli2017simultaneous}; it has enabled solution of various inclusion problems  \citep{selvadurai1981betti,selvadurai1982interaction,selvadurai2000application}; and extends to dynamics \citep{helmholtz1887ueber,lamb1887reciprocal},  as well as various additional fields, such as acoustics \citep{rayleigh1878treatise,howe1998acoustics}, optics \citep{helmholtz1856handbuch}, and fluids \citep{masoud2019reciprocal,daddi2018reciprocal,lorentz1896general}. 
The appeal of using the reciprocal theorem to understand the response at finite deformations, which can be exceedingly more difficult to capture experimentally, or to analytically resolve, is clear; even if for a limited class of boundary value problems. However, at the core of its proof  is the assumption of linear elasticity, which limits applications of the reciprocal theorem to infinitesimal deformations, or to small perturbations superposed upon an arbitrarily strained state of a hyperelastic material \citep{truesdell1963meaning,zorski1962equations}. Hence, it has not been previously used for finite deformations, and  cannot be generally extended to this range. Nevertheless, in this manuscript, we present a broad and general set of interchangeable boundary value problems, of practical significance,  for which the reciprocal theorem extends to large deformations in incompressible bodies.

In the next section, after describing the problem setting and constitutive assumptions, we will  present our theorem and will discuss relevant practical  applications. In Section \ref{Proof} we will provide a complete proof of the theorem and we will show that its solutions satisfy the Maxwell-Betti reciprocal theorem. We will conclude in Section \ref{conclusion}.

\section{The Theorem and its Applications}
\begin{figure}[h!]\centering
\includegraphics[width = 0.4\textwidth]{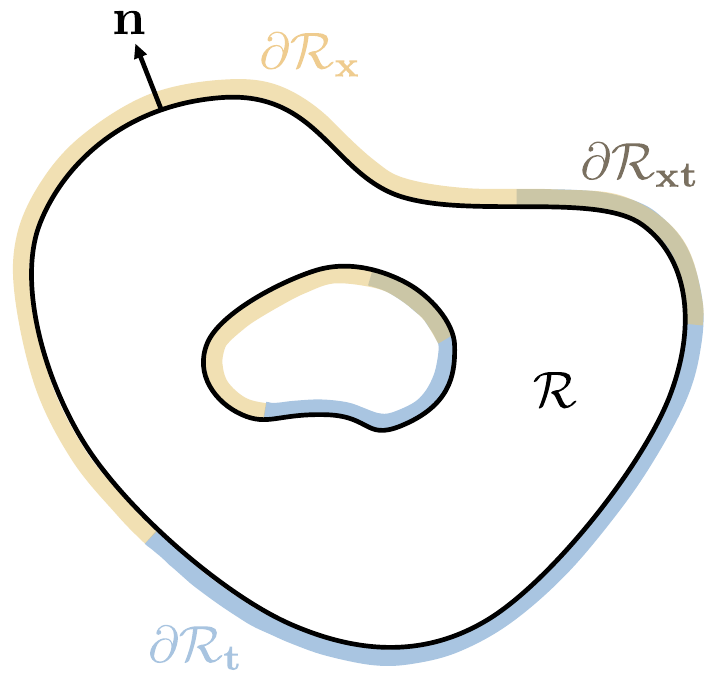}
\caption{Schematic illustration of the body region  $\mathcal{R}$ and its boundaries. The black lines define the boundaries of the non-simply connected body. Regions of the boundaries shaded  yellow and blue represent prescription of displacement ($\partial\mathcal{R}_{\bf x}$) and traction ($\partial\mathcal{R}_{\bf t}$) boundary conditions, respectively. The remaining shaded regions represent prescription of mixed boundary conditions  ($\partial\mathcal{R}_{\bf xt}$). }\label{illus}
\end{figure}

\noindent \textbf{{Problem Setting}.} Consider a body, $\mathcal{B}$, which occupies the regions  $\mathcal{R}$, and $\mathcal{R}^{\rm R}$ in its current and reference configurations, respectively\footnote{Throughout the manuscript will  use the superscript $(\ )^{\rm R}$, to distinguish reference quantities from current quantities, as we alternate between the two configurations for mathematical convenience.}. As illustrated in Figure  \ref{illus}, the body region need not be simply connected. A material point  is described in the reference configuration by its coordinates,
$\mathbf{X}$, and at time $t$ is mapped to its current spacial location, $\mathbf{x}$,  by the mapping $\mathbf{x}=\boldsymbol{\chi}(\mathbf{X},t)$. Accordingly, we can write the deformation gradient as $\mathbf{F}=\partial \boldsymbol{\chi}/\partial \mathbf{X}$. Additionally,  we parametrise the deformed  boundary surface of the body  $\partial\mathcal{R}$, by  $\mathbf{x}_b=\mathbf{x}(\mathbf{X}_b,t),$ where $\mathbf{X}_b$   is the collection of material points that define the undeformed boundary $\partial\mathcal{R}^{\rm R}$.

Let  $\boldsymbol{\sigma}(\mathbf{X},t)$ denote the Cauchy stress tensor field, and ${\mathbf{n}(\mathbf{X}_b,t)}$ the outward facing
normal to the boundary in the current configuration, we write  the surface traction in the current frame as $\mathbf{t}(\nten{X}_b,t)=\boldsymbol{\sigma}\mathbf{n}$.
As illustrated in Figure  \ref{illus}, we classify different regions on the boundary $\partial\mathcal{R}$ as follows:  $\partial\mathcal{R}_{\bf t}$ - denotes regions that are subjected to pure traction boundary conditions $\mathbf{t}=\mathbf{\bar t}$, where $\mathbf{\bar t}(\nten{X}_b,t)$ is the applied traction;    $\partial\mathcal{R}_{\bf x}$ - denotes regions that are subjected to pure displacement boundary conditions  $\mathbf{x}=\mathbf{\bar x}$, where $\mathbf{\bar x}(\mathbf{X}_b,t)$ is the prescribed  location; and   $\partial\mathcal{R}_{\bf xt}$ - denotes regions of mixed boundary conditions. 
 We also consider a body force field $\mathbf{b}(\mathbf{X},t)$. 

\vspace{1mm}

\noindent \textbf{Constitutive assumptions. }We restrict  our attention to a broad class of  incompressible bodies whose constitutive response can be described by a Helmholtz free energy function (per unit referential volume)\footnote{Note that restrictions on the specific form of $\psi(\mathbf{F})$, to ensure  frame indifference, can be found in  \cite{anand2020continuum}. Here we use the general representation for its concise  mathematical form.   }, $\psi(\mathbf{F})$. We denote the  constant mass density by  $\rho$, which implies $\det(\mathbf{F})=1$. Accordingly, the Cauchy stress tensor field takes the general form \cite{anand2020continuum} \begin{equation}\label{sig}
 \boldsymbol{\sigma}= \boldsymbol{\hat\sigma}(\mathbf{F})-q\mathbf{I} \qquad \text{in} \qquad \partial\mathcal{R}
\end{equation}where the \textit{hydrostatic pressure} field -  $q\mathbf{}(\mathbf{X},t)$ arises as a response to the incompressibility constraint and is constitutively indeterminate, and  $\boldsymbol{\hat\sigma}(\mathbf{F})$ is determined by the deformation gradient.  
\vspace{2mm}

\noindent \textbf{Auxiliary problem. }Next, consider an \textit{imaginary body}, $\mathcal{B}^{*}$, that is identical to $\mathcal{B}$ and occupies the same region $\mathcal{R}^{\rm R}$ in the reference configuration\footnote{Throughout the manuscript will  use the superimposed $(\ )^{\rm *}$, to denote quantities in the imaginary body of the auxiliary problem.}. It is subjected to the same body force  $\mathbf{b}$ and identical displacement boundary condition, but  to an alternate boundary traction,  that can be expressed as
\begin{equation}
\mathbf{\bar t}^*=\mathbf{\bar t}+p\mathbf{n}^* \qquad  \text{on} \qquad \partial\mathcal{R}_{\bf t}^*\cup\partial\mathcal{R}_{\bf xt}^*  \label{t*}
\end{equation}  where the constant $p$ can be thought of as an additional, externally applied,    \textit{hydrostatic pressure}. Note that in \cref{t*}, on the boundaries with mixed boundary conditions the extra pressure component is added to the prescribed traction $\mathbf{\bar t}^*$ only along the directions in which the traction $\mathbf{\bar t}$ is prescribed (see \cref{ti}). 
\vspace{2mm}

Provided the above kinematic description of the general problem setting and the restrictions on the constitutive response, we can now write our theorem as follows:

\vspace{2mm}

 \noindent\textbf{Theorem:} 
 
  \textit{The displacement field  $\mathbf{x}$, associated with  the boundary traction $\mathbf{\bar t}$, is identical to the displacement field   $\mathbf{x^{*}}$,  associated with the boundary traction $\mathbf{\bar t^{*}}$}.

\begin{figure}[H]\centering
\includegraphics[width = 0.9\textwidth]{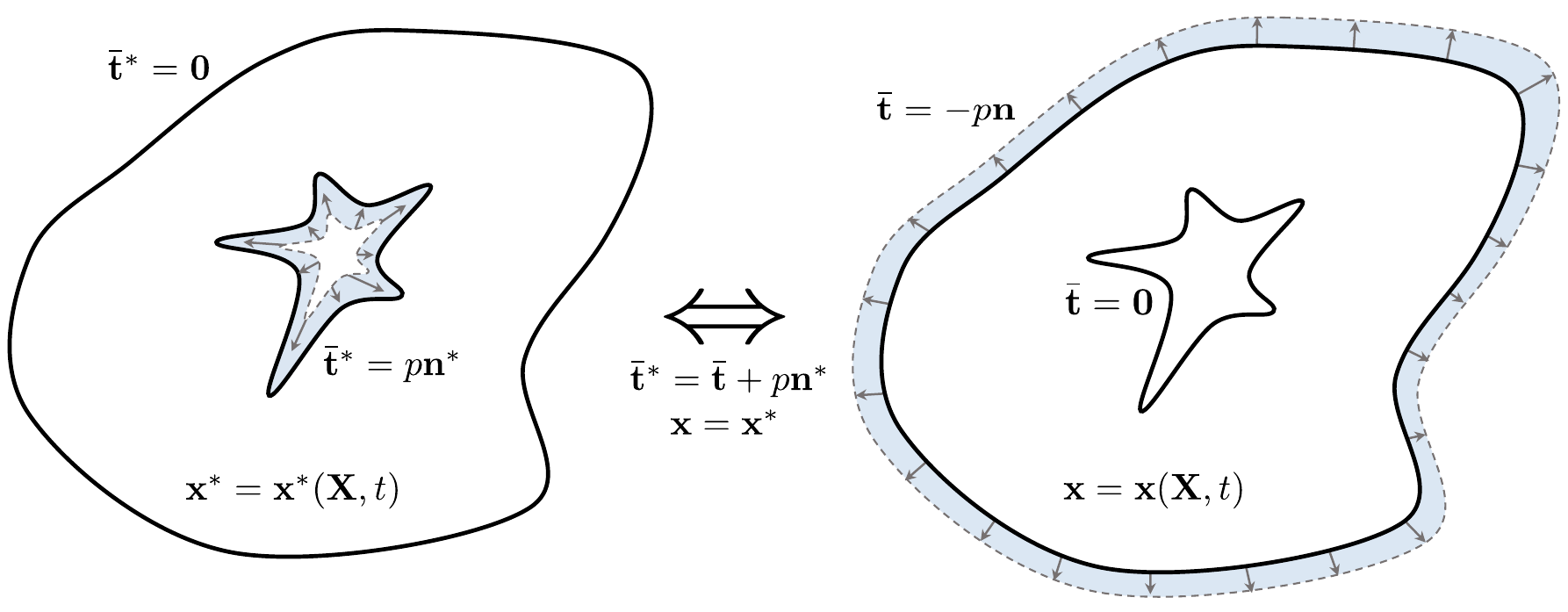}\vspace{-2mm}
\caption{Schematic illustration of an example application of the theorem for the expansion or contraction of an internal defect. On the left, we illustrate the auxiliary problem: An internal defect is subjected to a uniform hydrostatic pressure, $p$, while the external boundary of the body is stress free. The deformation that ensues from this loading, $\mathbf{x}^*$, is identical, to the deformation,  $\mathbf{x}$, that would be induced by application of a hydrostatic load of the same magnitude on the outer surface but with opposite sign. Note that the direction of the arrows in the illustration are representative of expansion process with $p<0$. The opposite sign, $p>0$ corresponds to contraction of the defect.      }\label{ex1}
\end{figure}

\vspace{2mm}
   \noindent\textbf{Examples.} 
Before providing a proof of the theorem we demonstrate some examples where its application allows for elegant non-trivial conclusions. First, consider the \textit{expansion or contraction of a   defect}, of arbitrary shape, within an incompressible body that is subjected to  hydrostatic load on its outer surface, as illustrated in Figure \ref{ex1}. 

\vspace{2mm}

  In an attempt to explain the limit of the resistance of a material to hydrostatic load, this fundamental problem has been studied extensively in the literature. The seminal study of  \cite{gent1959internal} reported an unusual rupture process in rubbers. That unstable rupture is now commonly referred to as cavitation \citep{horgan1995cavitation,knowles1965finite,ball1982discontinuous}, and is linked to the initiation of damage and fracture \citep{raayai2019intimate,quigley1994finite,ashby1989flow,lefevre2015cavitation}. However,  the  mechanical instability induced by application of external loads beyond a critical threshold can be an extremely fast and uncontrollable process;  attempts to experimentally study these internal ruptures   are thus challenging \citep{poulain2017damage,hutchens2016elastic}.  In contrast,  internal pressurization of a defect, by injection of an incompressible  fluid \citep{raayai2019volume,raayai2019capturing,chockalingam2021probing}, by phase separation \citep{kothari2020effect,style2018liquid}, or by the growth of an embedded inclusion \citep{li2022nonlinear,Zhange2107107118},  can allow for complete control over the expansion process, and is a promising avenue for measuring material properties and understanding the initiation of damage and fracture \citep{kim2020extreme,mijailovic2021localized,barney2020cavitation,franck2017microcavitation,estrada2018high}. In these settings however, the defect can have intricate shapes \citep{raayai2019intimate,yang2019hydraulic,milner2021multi}   and it is not obvious how the deformation field generated via internal pressurization translates to explain failure of the bulk material, as induced by application of external loads.

According to the above Theorem, regardless of the geometries of the body and the defect, application of internal hydrostatic pressure induces an identical deformation field, as  for the application of remote hydrostatic  tension of the same magnitude (Figure \ref{ex1}), thus supporting the use of  internal loading to study bulk failure in situations of external loading. Since the above theorem applies also for dynamic response, this may be particularly useful upon breakage of spherical  symmetry in Inertial Microcavitation Rheometry (IMR), where extreme dynamic loading is induced by spatially focused pulsed laser -- a tool that can explain onset of damage in biological tissue induced by extreme events, such as blast or impact \citep{PhysRevE.104.045108,yang2020extracting,franck_chapter}. Another instance where symmetry breaking has been observed is in the reverse scenario, where hydrostatic tension is applied at the cavity wall, and induces creasing \citep{milner2017creasing}. In this example, according to the above theorem, application of hydrostatic compression on the outer boundary will result in the same deformation field.   

We note  that the problem illustrated in Figure \ref{ex1}, can be further complicated\footnote{For this problem, $\{\partial \mathcal{R}_{\bf{x}}, \partial \mathcal{R}_{\bf{tx}}\} \in \varnothing$ (null set), $\partial \mathcal{R}_{\bf{t}} \equiv \partial \mathcal{R}$.} by regions of displacement boundary conditions, and mixed boundary conditions, without compromising the application of the theorem.  This can be particularly useful if the experimental specimen  is resting on a substrate or in a container while the internal expansion is performed.  At the limit, where the deformation field is symmetric (i.e. spherical or cylindrical), the above theorem has been analytically demonstrated \citep{knowles1965finite,cohen2015dynamic}.

 \begin{figure}[H]\centering
\includegraphics[width=\textwidth]{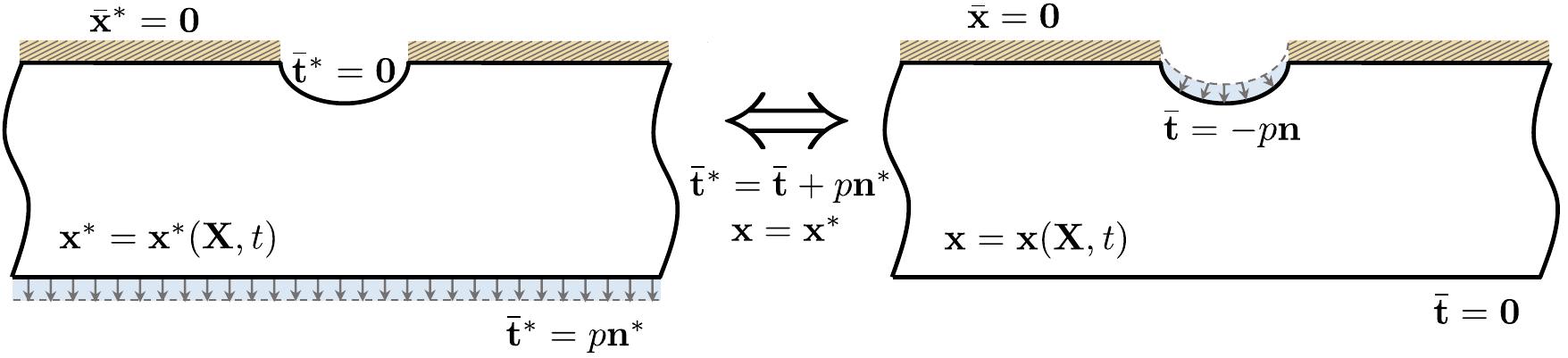}\vspace{-3mm}
\caption{Schematic illustration of an example application of the theorem for pressure indentation. On the left, we illustrate the auxiliary problem: A body that spans indefinitely in the plane is subjected to an applied uniform normal pressure, $p$ on its bottom surface and is fully constrained on its top surface, except for a small region that is traction free. The deformation that ensues from this loading, $\mathbf{x}^*$,  is identical to the deformation,  $\mathbf{x}$, that would be induced by application of  a pressure of the same magnitude but with opposite sign locally at the free region of the top interface while bottom surface is kept traction free.        }\label{ex2}
\end{figure}

The linear  reciprocal theorem is commonly employed for the interpretation of the force applied to an indentor in indentation experiments with different indentor shapes; analogously we can consider a setting of -- pressure indentation -- where a small region of an interface is detached from a rigid substrate and subjected to uniform hydrostatic pressure  \citep{wahdat2022pressurized}. This setting is relevant in \textit{blister tests}, for the measurement of interface properties \citep{chopin2008liquid,jensen1991blister,dannenberg1961measurement}, but can also emerge naturally by phase separation and condensation at an interface \citep{ma2020condensation,ma2019condensation}, or  by interfacial growth of biofilm --- a major cause of infections \citep{arciola2018implant,fortune2021biofilm}.   To illustrate this example,  we consider the body to span indefinitely in the plane and  assume that the bottom boundary is traction-free,    as illustrated in Figure \ref{ex2}. 

According to the above Theorem, application of a   normal tension of equal magnitude on the bottom surface would induce an identical deformation field as   the local indentation (Figure \ref{ex2}). This opens an avenue for development of simplified blister tests, whereby the load is applied externally on a larger area; it may thus be more  easily monitored, it can provide better visibility of the delamination process, and reduce issues related to compliance of the testing apparatus \citep{hohlfelder1994blister}. Formation of such interfacial cavities, by application of an external load, has been reported in an number of recent studies \citep{ringoot2021stick,kothari2020controlled,cohen2018competing}.  The additive effect of a remote pressure load on local processes can also aid in elucidating the build-up of pressure in interfacial condensates, and the role of applied loads in the development of chronic infections \citep{Zhange2107107118}.

In both of the above examples, alternating between the physical problem and the imaginary problem can provide an alternative and more convenient method for experimentation. Moreover, alternating between the  different sets of boundary conditions can have an advantage particularly in incompressible materials, where numerical implementation often employs an energy penalty on dilation, thus permitting small changes in volume. In certain settings, when the hydrostatic load is applied remotely, this can lead to significant computational errors. Finally, we would like to emphasize  that while we have  chosen these two specific examples for their simplicity, their modification to far more complex geometries (e.g. with numerous voids that may be interconnected) or loading conditions (e.g. not only hydrostatic loads,  but also alternating between traction, displacement, and mixed boundary conditions on any of the surfaces) does not compromise the applicability of the above theorem. 

Next, we proceed to provide the proof of the theorem.


\section{The proof}\label{Proof}
\noindent Let us begin by writing the boundary value problem that we wish to solve\footnote{The superimposed dot represents differentiation with respect to time.} \begin{align}
{\rm div}\boldsymbol{\sigma}+\mathbf{b}=\rho\ddot{\mathbf{x}} \qquad &\text{in} \qquad\mathcal{R} \label{em1}
\\
\mathbf{x}=\bar{\mathbf{x}}\ \ \qquad &\text{on} \qquad  \partial\mathcal{R}_{\bf x}\label{bc1}
\\
\mathbf{t}=   {\mathbf{\bar t}} \ \ \ \qquad &\text{on} \qquad  \partial\mathcal{R}_{\bf t}\label{bc2}
\\x_i=\bar x_i\quad  \text{or} \quad \mathbf{t}\cdot \mathbf{e}_i = \bar t_i     \qquad &\text{on} \qquad  \partial\mathcal{R}_{\bf xt} \quad \text{for} \quad i=1,2,3 \label{bc3}
\end{align}Here the first equation implies conservation of linear momentum and the latter three are  boundary conditions on the different boundary regions, where $\bar{\mathbf{x}}$ and $\bar{\mathbf{t}}$ are the prescribed displacements and boundary tractions, respectively. On regions of mixed boundary conditions, either displacement or traction components, $\bar x_i$ or $\bar t_i$, are prescribed along locally defined orthogonal directions, as represented by the unit vectors $ \mathbf{e}_i(\mathbf{X},t)$. Note that this general problem statement includes dynamic response, as reflected by the inertial term on the right-hand-side of \cref{em1}. 

Next, we use the definition in  \cref{t*} to write our auxiliary  problem statement as  \begin{align}
{\rm div}\boldsymbol{\sigma}^*+\mathbf{b}=\rho\ddot{\mathbf{x}}^* \qquad &\text{in} \qquad\mathcal{R}^* \label{em2}
\\
\mathbf{x}^*=\bar{\mathbf{x}} \ \ \ \  \qquad &\text{on} \qquad  \partial\mathcal{R}_{\bf x}^*\label{bc21}
\\
\mathbf{t}^{*} = {\mathbf{\bar t}}^*  \ \ \ \qquad &\text{on} \qquad  \partial\mathcal{R}_{\bf t}^*\label{bc22}
\\x^{*}_i=\bar x_i\quad  \text{or} \quad \mathbf{ t}^*\cdot \mathbf{e}_i = \bar t_i^*     \qquad &\text{on} \qquad  \partial\mathcal{R}_{\bf xt}^* \quad \text{for} \quad i=1,2,3 \label{bc23}
\end{align} 
   where \begin{equation}
\bar t_i^*=\bar t_i+p\mathbf{n}^*\cdot \mathbf{e}_i \label{ti}
\end{equation}  

\bigskip

\smallskip

\noindent{\bf{Proposition:}} \textit{If $\mathbf{x}^*(\mathbf{X},t)$ is a solution of the auxiliary boundary value problem \eqref{em2}-\eqref{ti}, then  
\begin{equation}\mathbf{x}(\mathbf{X},t)=\mathbf{x}^*(\mathbf{X},t)\label{post}\end{equation} is a solution of the boundary value problem \eqref{em1}-\eqref{bc3}.} 

\bigskip
If the above Proposition  is true, then the body and surface regions as well as  the surface normals  of both bodies in their deformed configurations are identical, namely \begin{equation} \mathcal{(R},\partial\mathcal{R}_{\bf x},\partial\mathcal{R}_{\bf t},\partial\mathcal{R}_{\bf xt})=(\mathcal{R}^{*},\partial\mathcal{R}^*_{\bf x},\partial\mathcal{R}_{\bf t}^{*},\partial\mathcal{R}_{\bf xt}^{*}), \quad \text{and}\quad \mathbf{n}=\mathbf{n}^* \label{identities}\end{equation}
This also implies that the deformation gradient fields are identical,  $\mathbf{F}^*(\mathbf{X},t)=\mathbf{F}(\mathbf{X},t)$, and thus that    $\boldsymbol{\hat\sigma}(\mathbf{F}^*)=\boldsymbol{\hat\sigma}(\mathbf{F})$. The stress field then follows  from   \cref{sig}  to write    
\begin{equation}
\boldsymbol{\sigma}=\boldsymbol{\sigma}^{*}- (q^{*}-q)\mathbf{I} 
\end{equation}where $q^{*}(\mathbf{X},t)$  and $q(\mathbf{X},t)$  are  hydrostatic pressure fields. Since $q^{*}-q$ is an arbitrary hydrostatic pressure field, consider   $q^{*}-q=p$ 
so that
\begin{equation}
\boldsymbol{\sigma}=\boldsymbol{\sigma}^*-p\mathbf{I} \label{post2}
\end{equation}

We can now substitute \cref{post,post2} in the boundary value problem \eqref{em1}-\eqref{bc3} while using \cref{identities} to write  \begin{align}
{\rm div}\boldsymbol{\sigma}^{*}-\cancelto{0}{{\rm div}(p\mathbf{I})}+\mathbf{b}=\rho\ddot{\mathbf{x}}^* \qquad &\text{in} \qquad\mathcal{R^{*}} \label{em3}
\\
\mathbf{x^{*}}=\bar{\mathbf{x}}\ \ \qquad &\text{on} \qquad  \partial\mathcal{R}^*_{\bf x}\label{bc31}
\\
\mathbf{\boldsymbol{\sigma}^*}\mathbf{n}^*={\mathbf{\bar{t}+p\mathbf{n}^*}} \ \qquad &\text{on} \qquad  \partial\mathcal{R}^*_{\bf t}\label{bc32}
\\x_i^*=\bar x_i\quad  \text{or} \quad (\boldsymbol{\sigma}^*\mathbf{n}^*)\cdot \mathbf{e}_i=\bar{t}_i+p\mathbf{n}^* \cdot \mathbf{e}_i    \qquad &\text{on} \qquad  \partial\mathcal{R}^*_{\bf xt} \quad \text{for} \quad i=1,2,3 \label{bc33}
\end{align} 
where we identify that the constant hydrostatic pressure vanishes from the equation of motion \eqref{em3}, and that the traction conditions  \eqref{bc32} and \eqref{bc33}, are identical to the applied tractions, $\mathbf{\bar{t}}^*$, and $\bar{t}_i^*$,    of the auxiliary boundary value problem \eqref{em2}-\eqref{ti}, as defined in \cref{t*}, and \cref{ti}. Accordingly, we have identically recovered  the auxiliary boundary value problem \eqref{em2}-\eqref{ti}. But, since  $\mathbf{x}^*$ is a solution of the auxiliary boundary value problem,    $\mathbf{x=\mathbf{x}^*}$  identically satisfies \eqref{em1}-\eqref{bc3} and our proposition is true. Consequently the proposed stress field $\boldsymbol{\sigma}$ in \cref{post2} is a solution of our boundary value problem.

It is important to mention that in nonlinear deformation, as generally considered here, there is no guarantee of uniqueness of the solution. Therefore, the solution obtained using this theorem is one potential solution. 

\subsection*{Identically satisfying the Maxwell-Betti theorem} 

The  elastodynamic reciprocal theorem, for infinitesimal deformations of an elastic body, as formulated by \cite{betti1872teoria} considers the displacement field $\mathbf{u}=\mathbf{x}-\mathbf{X}$, body force, $\mathbf{b}$, and applied tractions $\mathbf{t}$, and an auxiliary problem with auxiliary fields $\mathbf{x}^*,\mathbf{u}^*,\mathbf{b}^*, \mathbf{t}^*$, to write the analytical statement \citep{,love2013treatise}\footnote{Recall that in infinitesimal deformations,  there is no distinction between the reference and current frames. Hence, $\mathcal{R^{\rm R}}\equiv\mathcal{(R}= \mathcal{R}^*)$.}\begin{equation}
\iint\displaylimits_{\partial\mathcal{R}}\mathbf{t}\cdot\mathbf{u}^*{\rm d}A+\iiint\displaylimits_\mathcal{R}\left(\mathbf{b}-\rho \ddot{\mathbf{u}}\right)\cdot\mathbf{u}^*{\rm d}V=\iint\displaylimits_{\partial\mathcal{R^{}}}\mathbf{t^{*}}\cdot\mathbf{u}\ {\rm d}A+\iiint\displaylimits_\mathcal{R}\left(\mathbf{b}^{*}-\rho \ddot{\mathbf{u}}^*\right)\cdot\mathbf{u}{\ \rm d}V
\end{equation}An alternative form, which is more commonly used in fluid dynamics \citep{masoud2019reciprocal,leal1980particle}, considers reciprocity with respect to the velocity fields\footnote{Note that for fluids $\mathcal{R}$ corresponds to a control volume that is common to both problems. } $(\mathbf{v},{\mathbf{v}}^*)=(\dot{\mathbf{u}},\dot{\mathbf{u}}^*)$  \begin{equation}
\iint\displaylimits_{\partial\mathcal{R}}\mathbf{t}\cdot\mathbf{v}^*{\rm d}A+\iiint\displaylimits_\mathcal{R}\left(\mathbf{b}-\rho \dot{\mathbf{v}}\right)\cdot\mathbf{v}^*{\rm d}V=\iint\displaylimits_{\partial\mathcal{R^{}}}\mathbf{t^{*}}\cdot\mathbf{v}\ {\rm d}A+\iiint\displaylimits_\mathcal{R}\left(\mathbf{b}^{*}-\rho \dot{\mathbf{v}}^*\right)\cdot\mathbf{v}\ {\rm d}V\label{Navier}
\end{equation}While neither of the above analytical statements can be generally extended to finite elastic deformations, the latter form, which is based on the rate of change of internal energy in the system, rather than the total energy,  is more amenable for extension to special problems in finite deformations.       

In the present framework, we have limited our attention to a special class of problems, in which $\mathbf{b}^{*}=\mathbf{b}$, and   $\mathbf{t}^*=\mathbf{t}+p\mathbf{n}$. According to our theorem, which is formally stated in the proposition and in \cref{post}, this implies that $\mathbf{v}^*=\mathbf{v}$, and  $\dot{\mathbf{v}}^*=\dot{\mathbf{v}}$, with $\mathcal{R}=\mathcal{R}^*$. By substituting these identities into the above integral formula, we  notice that the volume integrals cancel, and we are left with \begin{equation}
\iint\displaylimits_{\partial\mathcal{R^{}}}p\mathbf{n}\cdot\mathbf{v}\ {\rm d}A=0\quad \Leftrightarrow \quad p\iiint\displaylimits_\mathcal{R}({\rm div~\mathbf{v})}\ {\rm  d}V=0
\end{equation}where we have used the \textit{divergence theorem} to arrive at the equivalent second equality. Finally, we notice that the second equality is identically satisfied for an incompressible material (i.e. if $J\equiv 1$ then ${\rm div~\mathbf{v}=0}$). Hence, the reciprocal theorem \eqref{Navier} is identically satisfied.

\section{Conclusion}\label{conclusion}
\noindent A reciprocal theorem that relates solutions of a specific class of large deformation boundary value problems for incompressible bodies is presented and the complete proof is provided. In essence, the theorem states that the addition of uniform normal pressure to the boundary traction 
does not change the deformation field solution to the problem and the stress field solution differs only by a uniform hydrostatic component whose magnitude is equal to the magnitude of the added pressure. Although limited to incompressible bodies, this theorem is relevant in various modern applications in mechanics of soft materials where the assumption of incompressibility is frequently employed; some examples that lead to elegant non-trivial conclusions are discussed. Finally, it is shown that the solution fields of the  presented theorem identically satisfy the classical Maxwell-Betti theorem. Future work could reveal potential applications of the theorem not identified here. In addition, generalization of the theorem to a broader class of incompressible constitutive response, such as rate dependence or higher-order elasticity theories might be possible and is left for future work.

\section*{Acknowledgements}
\noindent The authors wish to acknowledge the support of: the Army Research Office, United States of
America and Dr. Ralph A. Anthenien, Programme Manager, under award number W911NF-19-1-0275; and
the Office of Naval Research, United States of America and Dr. Timothy B. Bentley, Programme Manager,
under award number N00014-20-1-2561. The authors are grateful to Rohan Abeyaratne for insightful conversations.

\bibliographystyle{elsarticle-harv}
\bibliography{mybibfile} 

\end{document}

%% file: settings.tex
\usepackage{amsmath}

\usepackage{amssymb}
\usepackage{subfiles} 
\usepackage{subcaption}
\usepackage{multicol}
\usepackage{bm}
\usepackage{graphicx}
\usepackage[export]{adjustbox}
\usepackage[font=footnotesize]{caption}
\usepackage{float} 
\usepackage{placeins}
\usepackage[dvipsnames]{xcolor}

\definecolor{red}{rgb}{1,0,0}
\definecolor{black}{rgb}{0,0,0}
\definecolor{codegreen}{rgb}{0,0.6,0}
\definecolor{codegray}{rgb}{0.5,0.5,0.5}
\definecolor{codepurple}{rgb}{0.65,0,0.82}
\definecolor{backcolour}{rgb}{0.97,0.97,0.97}
\definecolor{codeorange}{rgb}{1,0.31,0}
\usepackage{longtable}

\usepackage{xparse}
        \ExplSyntaxOn
        \NewDocumentCommand{\convertto}{mm}
         {  \texttt{#2~=~\fp_to_decimal:n { (#2)/(1#1) }#1} }
         \ExplSyntaxOff

\usepackage[per-mode=symbol]{siunitx}